\RequirePackage{fix-cm}
\documentclass[pdftex,twocolumn,epjc3]{svjour3}  

\RequirePackage[T1]{fontenc}

\usepackage[vcentermath]{youngtab}
\usepackage{amssymb}
\usepackage{amsmath}
\usepackage{mathtools}
\usepackage{slashed}
\usepackage{graphicx}
\usepackage{multirow}
\usepackage{hyperref}
\usepackage{longtable}
\usepackage{xcolor}
\usepackage{float}
\usepackage{bm}

\newcommand{\bra}[1]{\langle #1|}
\newcommand{\ket}[1]{|#1\rangle}

\def\w2{\tilde w^2}
\def\ws2{1}

\allowdisplaybreaks
\smartqed  
\journalname{Eur. Phys. J. A}
\begin{document}

\title{Estimation of coupling constants for D-meson, charmed, and light baryons in effective Lagrangian approach and quark model 
}


\author{Thanat Sangkhakrit\thanksref{e1,SUT}
        \and
        Attaphon Kaewsnod\thanksref{SUT}
        \and
        Warintorn Sreethawong\thanksref{SUT}
        \and
        Thananuwat~Suyuporn\thanksref{SUT}
        \and
        Nopmanee Supanam\thanksref{SWU}
        \and
        Daris Samart\thanksref{KKU}
        \and \\
        Yupeng Yan\thanksref{e2,SUT} 
}

\thankstext{e1}{e-mail: tanattosan@outlook.jp}
\thankstext{e2}{e-mail: yupeng@g.sut.ac.th}


\institute{School of Physics and Center of Excellence in High Energy Physics and Astrophysics,
	Suranaree University of Technology, Nakhon Ratchasima 30000, Thailand \label{SUT}
           \and
    Department of Physics, Srinakharinwirot University, Bangkok, 10110, Thailand \label{SWU}
           \and
    Department of Physics, Faculty of Science, Khon Kaen University, 123 Mitraphap Rd., Khon Kaen, 40002, Thailand \label{KKU}
}

\date{Received: date / Accepted: date}

\maketitle

\begin{abstract}
We estimate coupling constants for effective Lagrangians of $D$-meson, charmed, and light baryons from charmed baryon decay processes.
First, we calculate decay widths for the processes $\Lambda_{c} \to DN$, $\Sigma_{c} \to D \Delta$, $\Sigma_{c} \to D^{*} \Delta$, and $\Lambda_{c} \to D^{*}N$ in effective Lagrangian method and quark model picture with $^{3}P_{0}$ model.
By employing the coupling constants for $D^{*} \Lambda_{c} N$ interaction from several literatures, the strength parameter $\lambda$ for $^{3}P_{0}$ quark model is fixed in the decay process $\Lambda_{c} \to D^{*}N$.
Then, the coupling constants for the effective Lagrangians of $D\Lambda_{c}N$, $D\Sigma_{c}\Delta$, and $D^{*}\Sigma_{c}\Delta$ interactions are estimated in the decay channels $\Lambda_{c} \to DN$, $\Sigma_{c} \to D \Delta$, and $\Sigma_{c} \to D^{*} \Delta$, respectively.
These coupling constants will be useful for further studies of charm hadrons.

\end{abstract}

\section{Introduction}
\label{sec-1}


Physics of charm hadrons has been one of the main subjects in hadron physics since the first observations of $J/\psi$ meson in 1974 \cite{Augustin:1974xw,Aubert:1974js} and of charmed baryons ($\Lambda_{c},\Sigma_{c}$) in 1975 \cite{Cazzoli:1975et}.
Ever since, experimental observations for various exotic hadrons have been reported by Belle, BABAR, BESIII, and LHCb Collaborations \cite{Choi:2003ue,Aubert:2004ns,Aubert:2005rm,Abe:2007jna,Choi:2007wga,Hosaka:2016pey,Ablikim:2013mio,Liu:2013dau,Ablikim:2013wzq,Aaij:2013zoa,Aaij:2014jqa}, and theoretical studies have been carried out in a variety of models \cite{Micu:1968mk,Godfrey:1985xj,Maiani:2004vq,Ebert:2005nc,Limphirat:2010zz,Xu:2020ppr,Gupta:1994mw,Ebert:1997nk,Glozman:2003bt,Nowak:2003ra,AlFiky:2005jd,Liu:2006jx,Brambilla:1999xf,Braguta:2005kr,Isgur:1984bm,Chen:2000ej,Okamoto:2001jb,Liao:2002rj,McNeile:2002az,Chiu:2005ey,Chiu:2006hd} (See Refs.\cite{Swanson:2006st,Brambilla:2010cs} for reviews).
While charmed mesons have been extensively investigated, the properties of charmed baryons are less known since they have not yet been explored in the same detail. 
Proposals for charmed baryons study have been planned at future experiments at $\bar{\text{P}}$ANDA \cite{Wiedner:2011mf} and J-PARC \cite{e50} and the facilities are now under preparation. Thus, theoretical study of their production is important. The production rate of charmed baryons will be crucial to guide and assess these experimental plans. Moreover, their production mechanism provides not only the information of their internal structure and non-perturbative QCD dynamics, but also the role of chiral and heavy quark symmetries in heavy-light quark systems. 

One of the most important ingredients for calculation of charmed baryon production rate is the coupling constants. So far, it is not possible to determine the coupling constants for the effective Lagrangians of $D$-meson, charmed, and light baryons directly from the existing data.
Therefore, several methods have been used to extract the coupling constants for various charmed baryon interaction vertices.
In Refs.\cite{Kim:2015ita}, coupling constants for strange hadrons derived from Nijmegen potential are employed to study charmed productions from pion-proton collisions.
The coupling constants in these studies are of the same order as those in Ref.\cite{Sangkhakrit:2020wyi}, where they are determined from the $SU(3)$ symmetry relations and from the fit with the observed data for strangeness productions.
In Ref.\cite{Khodjamirian:2011sp}, coupling constants derived from light cone QCD sum-rules are employed to predict charmed hadron production cross sections at $\bar{\text{P}}$ANDA. 
On the other hand, the coupling constants from the $SU(4)$ symmetry are utilized to study the production of charmed baryons from proton-antiproton collisions in Refs.\cite{Haidenbauer:1991kt,Titov:2008yf,Shyam:2014dia,Haidenbauer:2016pva}.
From the previous studies, different sets of coupling constants result in discrepancies in the predicted charm production rates.

In this study, we estimate the coupling constants for the effective Lagrangians of $D$-meson, charmed, and light baryons from the decay widths of $\Lambda_{c}$ and $\Sigma_{c}$ baryons.
Firstly, the decay widths for the processes $\Lambda_{c} \to DN$, $\Lambda_{c} \to D^{*}N$, $\Sigma_{c} \to D \Delta$, and $\Sigma_{c} \to D^{*} \Delta$ are computed in effective Lagrangian method and quark model picture with $^{3}P_{0}$ model.
Then, the strength parameter $\lambda$ of the $^{3}P_{0}$ quark model is fixed from the decay channel $\Lambda_{c} \to D^{*}  N$, where the coupling constants for $D^{*}\Lambda_{c}N$ vertex from several literatures are used as inputs.
The coupling constants for the vertices $D\Lambda_{c}N$, $D\Sigma_{c}\Delta$, and $D^{*}\Sigma_{c}\Delta$ are consequently estimated in the corresponding decay channels.
In this work, we present four sets of the coupling constants for $D\Lambda_{c}N$, $D\Sigma_{c}\Delta$, and $D^{*}\Sigma_{c}\Delta$ interactions.

The content of this paper is organised as follows.
In Sec.\ref{sec-2}, we compute the decay widths of charmed baryons in effective Lagrangian method.
In Sec.\ref{sec-3}, the calculations of the same decay processes from Sec.\ref{sec-2} are performed in quark model with $^{3}P_{0}$ model.
Then, the estimation of the coupling constants from the two models is presented in Sec.\ref{sec-4}.
Finally, the summary of this study is given in Sec.\ref{sec-5}.

\section{Effective Lagrangian method}
\label{sec-2}

\begin{figure}[h]
	\centering
	\includegraphics[scale=1.0]{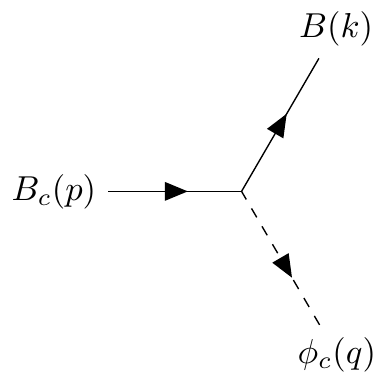}
	\caption{Feynman diagram for the decay process $B_{c} \to B \phi_{c}$.}
	\label{FM}
\end{figure}

In this section, we calculate decay widths of charmed baryons in effective Lagrangian method.
The decay of an initial charmed baryon $B_{c}$ into an outgoing light baryon $B$ and a charmed meson $\phi_{c}$ is displayed by the diagram in Fig.\ref{FM}.
Here, the momentum of the initial charmed baryon $\left( \Lambda_{c}(2286) \text{ or } \Sigma_{c}(2455) \right)$ is denoted by $p$, while $k$ and $q$ are those of the outgoing light baryon $\left( N(939) \text{ or } \Delta(1232) \right)$ and charmed meson $\left( D(1868) \text{ or } \right.$ $\left. D^{*}(2009) \right)$ respectively.

The effective Lagrangians for $D B_{c} B$ interaction vertices are given by
\begin{align}
	\mathcal{L}_{D\Lambda_{c}N}^{(A)} &= -\frac{g_{0}}{m_{D}} \bar{N}\gamma^{\mu}\gamma_{5}\Lambda_{c} \partial_{\mu}D, \label{LD1} \\
	\mathcal{L}_{D\Lambda_{c}N}^{(P)} &= g_{1} \bar{N} i\gamma_{5} \Lambda_{c} D, \label{LD2} \\
	\mathcal{L}_{D\Sigma_{c}\Delta} &= \frac{g_{2}}{m_{D}} \bar{\Delta}^{\mu} \boldsymbol{\Sigma}_{c} \cdot \boldsymbol{T} \partial_{\mu} D.
	\label{LD3}
\end{align}
For $D^{*} B_{c} B$ interaction vertices, we introduce the following Lagrangians
\begin{align}
	\mathcal{L}_{D^{*}\Lambda_{c}N} &= f_{0} \bar{N} \gamma^{\mu} \Lambda_{c} D_{\mu}
	+ \dfrac{h_{0}}{m_{D}} \bar{N} \sigma^{\mu \nu} \Lambda_{c} \partial_{\nu} D_{\mu}, \label{LDs1} \\
	\mathcal{L}_{D^{*}\Sigma_{c}\Delta} &= f_{1} \bar{\Delta}^{\mu} \gamma_{5} \boldsymbol{\Sigma}_{c} \cdot \boldsymbol{T} D_{\mu},
	\label{LDs2}
\end{align}
where $m_{D}$ corresponds to the approximate mass of the pseudoscalar $D$-meson and $\boldsymbol{T} = (T_{1},T_{2},T_{3})$ represents the isospin transition matrices operating on the isospin states of $\Delta$ and $D$ (or $D^{*}$).

By employing the Lagrangians in Eqs.(\ref{LD1})-(\ref{LDs2}), Feynman amplitudes for the decay processes $\Lambda_{c} \to D N$, $\Lambda_{c} \to D^{*}N$, $\Sigma_{c} \to D\Delta$, and $\Sigma_{c} \to D^{*}\Delta$ are written as
\begin{align}
	\mathcal{M}^{(A)}_{\Lambda_{c} \to DN} &= \dfrac{g_{0}}{m_{D}} \bar{u}_{N}\left(k,s^{\prime}\right) \slashed{q} \gamma_{5} u_{\Lambda_{c}}\left(p,s\right), \\
	\mathcal{M}^{(p)}_{\Lambda_{c} \to DN} &= - g_{1} \bar{u}_{N}\left(k,s^{\prime}\right) \gamma_{5} u_{\Lambda_{c}}\left(p,s\right), \\
	\mathcal{M}_{\Lambda_{c} \to D^{*}N} &= i f_{0} \bar{u}_{N}\left(k,s^{\prime}\right) \Gamma^{\mu} u_{\Lambda_{c}}\left(p,s\right)\epsilon^{*}_{\mu} \left(q,s''\right), \\
	\mathcal{M}_{\Sigma_{c} \to D\Delta} &= -\dfrac{g_{2}}{m_{D}} q_{\mu}\bar{u}^{\mu}_{\Delta}\left(k,s^{\prime}\right) u_{\Sigma_{c}}\left(p,s\right), \\
	\mathcal{M}_{\Sigma_{c} \to D^{*}\Delta} &= i f_{1} \bar{u}^{\mu}_{\Delta}\left(k,s^{\prime}\right) \gamma_{5} u_{\Sigma_{c}}\left(p,s\right) \epsilon^{*}_{\mu} \left(q,s''\right),
	\label{EFL}
\end{align}
where
\begin{equation}
	\Gamma^{\mu} = \left[ \gamma^{\mu} + \dfrac{i}{m_{D}}\left(\dfrac{h_{0}}{f_{0}}\right) \sigma^{\mu \nu} q_{\nu} \right].
\end{equation}
The spin projections of the initial charmed baryon, outgoing light baryon and $D^{*}$-meson are respectively denoted by $s$, $s^{\prime}$, and $s''$.
The decay width of the initial charmed baryon $B_{c}$ is then computed from
\begin{equation}
	\Gamma_{\text{EFT}} =  \dfrac{1}{32 \pi^{2}} \frac{\left|\vec{q}\right|}{m_{B_{c}}}\int \left< \left|\mathcal{M}\right|^{2} \right> d\Omega,
	\label{EFTW}
\end{equation}
where
\begin{equation}
	\left< \left|\mathcal{M}\right|^{2} \right> =
	\begin{cases}
		\dfrac{1}{2} \sum_{s^{\prime}}\left|\mathcal{M}\right|^{2} & \text{if } \phi_{c} = D, \\
		\vspace{0.5 pt} \\
		\dfrac{1}{2} \sum_{s^{\prime},s''}\left|\mathcal{M}\right|^{2} & \text{if } \phi_{c} = D^{*}.
	\end{cases}
\end{equation}
The mass of the initial charmed baryon and the magnitude of outgoing 3-momentum in the center of mass frame of the initial charmed baryon are denoted by $m_{B_{c}}$ and $\left|\vec{q}\right|$.

By expanding the decay width with respect to the outgoing 3-momentum $q$ near the threshold, the following expressions for the decay widths are obtained
\begin{align}
	\Gamma^{(A)}_{\Lambda_{c} \to DN} &= \frac{g_{0}^{2} \left( m_{D}+2m_{N} \right)^{2}}{8 \pi m_{D}^{2} m_{N} m_{\Lambda_{c}}} q^{3} \label{W1}, \\
	\Gamma^{(P)}_{\Lambda_{c} \to DN} &= \frac{g_{1}^{2}}{8 \pi m_{N} m_{\Lambda_{c}}} q^{3} \label{W2}, \\
	\Gamma_{\Lambda_{c} \to D^{*}N} &= \frac{\mathcal{A}}{8 \pi m_{D}^{2} m_{D^{*}}^{2} m_{N} m_{\Lambda_{c}}}q^{3} \label{W3}, \\
	\Gamma_{\Sigma_{c} \to D\Delta} &= \frac{g_{2}^{2} \left( m_{D}+m_{\Delta} \right)^{2}}{3 \pi m_{D}^{2} m_{\Delta} m_{\Sigma_{c}}}q^{3} \label{W4}, \\
	\Gamma_{\Sigma_{c} \to D^{*}\Delta} &= \frac{f_{1}^{2}}{4 \pi m_{\Delta} m_{\Sigma_{c}}} q^{3},
	\label{W5}
\end{align}
where
\begin{align}
	\mathcal{A} = &f_{0}^{2} \left( 3m_{D^{*}}^{2}m_{D}^{2} + 4m_{D^{*}} m_{N} m_{D}^{2} + 4 m_{N}^{2} m_{D}^{2} \right) \nonumber \\
	&- 6f_{0}h_{0} \left( m_{D^{*}}^{3}m_{D} + 2m_{D^{*}}^{2}m_{N}m_{D} \right) \nonumber \\
	&+ h_{0}^{2} \left( 3 m_{D^{*}}^{4} + 8 m_{D^{*}}^{3} m_{N} + 8 m_{N}^{2} m_{D^{*}}^{2} \right).
\end{align}
We note that the decay widths in Eqs.(\ref{W1})-(\ref{W5}) hold for real and imaginary outgoing momenta.

\section{$^{3}P_{0}$ quark model}
\label{sec-3}

\begin{figure}[H]
	\centering
	\includegraphics[width=0.4\textwidth]{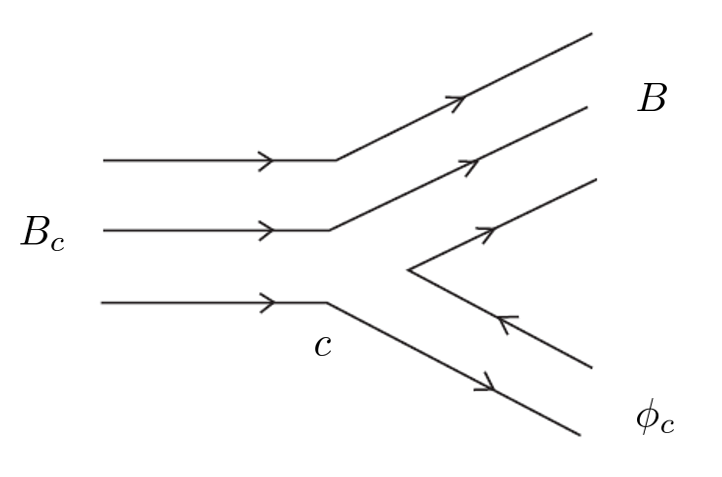}
	\caption{Schematic diagram for the decay process $B_{c} \to B\phi_{c}$ in $^{3}P_{0}$ quark model. The bottom quark line is that of charm quark while the rest are those of $u$ and $d$ quarks.}
	\label{fig:B-to-BM}
\end{figure}

In this section, decay widths of the same decay processes as in Sec.\ref{sec-2} are calculated in a quark model picture with the $^{3}P_{0}$ model.
The corresponding diagram is displayed in Fig. \ref{fig:B-to-BM}.
Here, the decay process $B_c\to B\phi_c$ may arise from the $qq$ and $c$ of the initial state $B_c$ which are directly dressed by two additional quark-antiquark pair pumped out of the vacuum to form $B$ and $\phi_c$ in the final state.

The transition amplitude derived in the $^3P_0$ model is written as
\begin{align}
	T=\bra{B\phi_c}V_{q\bar q}\ket{B_{c}},
\end{align}
where $V_{q\bar q}$ corresponds to the effective quark-antiquark vertex.
The $^3P_0$ model defines the quantum states of quark-antiquark pair that are destroyed into or created from vacuum ($^3P_0$, isospin $I=0$, and color singlet).
The effective quark-antiquark vertex in the $^3P_0$ model is defined according to Refs. \cite{Yan:2004jg,Kittimanapun:2008wg}:
\begin{align}
	V^{ij}_{q\bar q}=&\ \lambda\,\vec{\sigma}_{ij}\cdot(\vec{p}_i-\vec{p}_j)\hat F_{ij}\hat C_{ij}\delta(\vec{p}_i+\vec{p}_j)\nonumber\\
	=&\ \lambda\sum_\mu\sqrt{\frac{4\pi}{3}}(-1)^\mu\sigma^\mu_{ij}Y_{1\mu}(\vec{p}_i-\vec{p}_j)\hat F_{ij}\hat C_{ij}\delta(\vec{p}_i+\vec{p}_j) \nonumber\\
\end{align}
where the parameter $\lambda$ denotes the effective coupling strength of the $^3P_0$ vertex.
The spin operator that creates (or annihilates) the spin-1 $q\bar{q}$ pair is denoted by $\sigma^\mu_{ij}$ and $Y_{1\mu}(\vec{p})$ corresponds to the spherical harmonics in the momentum space.
The flavor and color unit operators are denoted by $\hat F_{ij}$ and $\hat C_{ij}$.

In this work, the baryon and meson spatial wave functions are approximated with the Gaussian form \cite{Faessler:2010zzc}.
The flavor and spin parts are constructed in the framework of the $SU(2)$ flavor and $SU(2)$ spin symmetries.
The transition amplitude is obtained as
\begin{equation}
	T = \lambda\sqrt{\dfrac{4\pi}{3}}C_i\ fe^{-Qq^2}C(S_is_i;1\mu;S_f,s_i+\mu),
	\label{QMA}
\end{equation}
with
\begin{align}
	f=&-\dfrac{6\sqrt{3}a^3b^{3/2}(b^2m_r+2a^2(1+m_r))|\vec q|}{(3a^2+b^2)^{5/2}(1+m_r)\pi^{3/4}},\nonumber\\\nonumber\\
	Q=&\dfrac{a^2(3a^2(1+m_r)^2+b^2(5-2m_r+2m_r^2))}{6(3a^2+b^2)(1+m_r)^2},\nonumber\\
	\nonumber \\
	C_i=&\frac{2}{\sqrt{3}}\left(\frac{1}{\sqrt{2}}\right)^{\delta_{S',\frac{1}{2}}}\nonumber\\
	&\sqrt{(2S'+1)(2S''+1)(2S_i+1)(3)}\nonumber\\
	&\sqrt{(2T'+1)(2T''+1)(2T_i+1)(1)}\nonumber\\
	&\begin{Bmatrix}
		T_{i} & (6) & S'\\
		(8)  & (7) & S''\\
		S_{i}  & 1 & S_f\\
	\end{Bmatrix}
	\begin{Bmatrix}
		T_{i} & (6) & T'\\
		0  & (7) & T''\\
		T_{i}  & 0 & T_f\\
	\end{Bmatrix},
\end{align}
where $\left(S_{i}, T_{i}\right)$, $\left(S^{\prime}, T^{\prime}\right)$, and $\left(S^{\prime\prime}, T^{\prime\prime}\right)$ denote the spin-isospin of the states $B_{c}$, $B$, and $\phi_{c}$, respectively.
The spin $S_{f}$ and isospin $T_{f}$ are defined by $S_f=S'\otimes S''$ and $T_f=T'\otimes T''$.
The spin projections of the $q\bar q$ pair in the $^3P_0$ model and the initial charmed baryon $B_{c}$ are denoted by $\mu$ and $s_{i}$.
$C$ is the Clebsch-Gordan coefficient.
The parameter $m_r=m_q/m_Q$ is the ratio between the light quark mass $m_{q}$ and heavy quark mass $m_{Q}$. The value of $m_{r}$ in this study is $300/1270$.
$\delta$ is the Kronecker delta and the brackets $\{\ \}$ in $C_i$ are the 9-j symbols.
The flavor-spin-color factors $C_i$ for the decay processes in this study are summarized in TABLE \ref{Tab:QM}.
The baryon and meson length parameters $a$ and $b$ are respectively 3.0 GeV$^{-1}$ and 2.28 GeV$^{-1}$ \cite{Isgur:1979be,Sreethawong:2014jra,Dover:1992vj,Muhm:1996tx,Limphirat:2013jga}.

\begin{table*}[t]
	\centering
	\begin{tabular}{lclc}
		\hline\hline
		\multirow{2}{*}{Processes}               & \multicolumn{3}{c}{$C_i$}                                    \\ \cline{2-4}
		& $S_{f}=1/2$                        & \hspace{1em}  & $S_{f}=3/2$              \\ \hline
		$\Lambda_c(2286)\rightarrow ND$        & $\frac{1}{\sqrt{2}}$             &  &                        \\ \hline
		$\Lambda_c(2286)\rightarrow ND^*$      & $-\frac{1}{\sqrt{6}}$            &  & $\sqrt{\frac{2}{3}}$   \\ \hline
		$\Sigma_c(2455)\rightarrow \Delta D$   & $-\frac{2}{3}\sqrt{\frac{2}{3}}$ &  &                        \\ \hline
		$\Sigma_c(2455)\rightarrow \Delta D^*$ & $\frac{4}{9}$                    &  & $\frac{2\sqrt{10}}{9}$ \\ \hline\hline
	\end{tabular}
	\caption{The flavor-spin-color factors $C_i$ corresponding to the decay processes $B_c 	\to B\phi_c$.}
	\label{Tab:QM}
\end{table*}
\begin{table*}[t]
	\begin{center}
		\begin{tabular}{|c|c|c|c|c|c|c|}
			\cline{1-7}
			\multirow{2}{*}{Ref.} & \multicolumn{2}{|c|}{Input} & \multicolumn{4}{|c|}{Results}
			\\
			\cline{2-7}
			& $f_{D^{*}\Lambda_{c}N}$ & $h_{D^{*}\Lambda_{c}N}$ & $g^{(P)}_{D\Lambda_{c}N}$ & $g^{(A)}_{D\Lambda_{c}N}$ & $g_{D\Sigma_{c}\Delta}$ & $f_{D^{*}\Sigma_{c}\Delta}$ \\
			\cline{1-7}
			\cite{Kim:2015ita} & -4.26 & -12.4 & 17.57 & 15.12 (13.4) & 9.28 & 26.05  \\
			\cite{Sangkhakrit:2020wyi} & -5.11 & -10.4 & 16.65  & 14.33 (13.5) & 8.79 & 24.68 \\
			\cite{Khodjamirian:2011sp} & 5.8 & 3.6 & 11.1 (10.7) & 9.55 & 5.86 & 16.45 \\
			\cite{Titov:2008yf}	& -5.18 & -14.4 & 20.66 & 17.78 & 10.9 & 30.62  \\
			\cline{1-7}
		\end{tabular}
	\end{center}
	\caption{Coupling constants of $D$-meson, charmed, and light baryons from our estimation. The numbers in the brackets denote the magnitudes of the original coupling constants used in the cited literatures (if available).}
	\label{coupling}
\end{table*}

The decay width of the charmed baryon $B_{c}$ is calculated from
\begin{align}
	\Gamma_{\text{QM}} &= \dfrac{2\pi E'E''\left|\vec{q}\right|}{m_{B_{c}}(2S_i + 1)} \displaystyle\sum_{s_i,\mu,S_f} \left|T\right|^{2},
	\label{QMW}
\end{align}
where $E'$ and $E''$ denote energies of the outgoing light baryon $B$ and charmed meson $\phi_c$ while $\left|\vec{q}\right|$ and $m_{B_{c}}$ are similar to those in Eq.\ref{EFTW}.

\section{Estimation of coupling constants with charmed baryon}
\label{sec-4}

In this section, we estimate the coupling constants for the effective Lagrangians in Eqs.(\ref{LD1})-(\ref{LDs2}) from the decay widths calculated in Sec.\ref{sec-2} and Sec.\ref{sec-3}.
Considering that the decay width formulas in Eqs.(\ref{W1})-(\ref{W4}) and Eq.(\ref{QMW}) hold for both the real and imaginary values of the outgoing momentum $q$, one may estimate the coupling constants by applying the near threshold off-shell decay processes of $\Lambda_{c}$ and $\Sigma_{c}$ baryons under consideration. In the low $q$ region, one requires
\begin{equation}
	\Gamma_{EFT} = \Gamma_{QM}.
	\label{solve}
\end{equation}
For comparison, we employ as inputs four different sets of the coupling constants $f_{D^{*}\Lambda_{c}N}$ and $h_{D^{*}\Lambda_{c}N}$ from Refs.\cite{Titov:2008yf,Kim:2015ita,Sangkhakrit:2020wyi,Khodjamirian:2011sp} for the decay process $\Lambda_{c} \to D^{*}N$. From Eq.(\ref{solve}), we fix the $^3P_0$ strength parameter $\lambda$ in Eq.(\ref{QMA}) for each input set and then use its value to estimate the coupling constants $g_{D\Lambda_{c}N}$, $g_{D\Sigma_{c}\Delta}$, and $f_{D^{*}\Sigma_{c}\Delta}$ of the effective Lagrangians.
In our case, we assume that all coupling constants are positive and they are displayed in TABLE.\ref{coupling}.
Note that the expressions for the coupling constants resulted from Eq. (\ref{solve}) are independent of the corresponding initial masses.

Here, we have used $g^{(P)}_{D\Lambda_{c}N} = g_{1}$, $g_{D\Sigma_{c}\Delta} = g_{2}$, and $f_{D^{*}\Sigma_{c}\Delta} = f_{1}$.
The coupling constants $f_{D^{*}\Lambda_{c}N}$, $h_{D^{*}\Lambda_{c}N}$, and $g^{(A)}_{D\Lambda_{c}N}$ are obtained by rescaling the coupling constants in Eq.(\ref{LD1}) and Eq.(\ref{LDs1}) to those in Refs.\cite{Kim:2015ita,Sangkhakrit:2020wyi,Titov:2008yf}.
From our estimation, we have found that the magnitudes of the coupling constants $g^{(P)}_{D\Lambda_{c}N}$ and $g^{(A)}_{D\Lambda_{c}N}$ agree with those in Ref.\cite{Kim:2015ita,Sangkhakrit:2020wyi,Khodjamirian:2011sp}.
In Ref.\cite{Titov:2008yf}, this interaction vertex is neglected since the vector dominance has been assumed.
As the original values of $g_{D\Sigma_{c}\Delta}$ and $f_{D^{*}\Sigma_{c}\Delta}$ are not presented anywhere, we only display the results from our estimation.

\section{Summary and conclusion}
\label{sec-4}

In this study, we have estimated the coupling constants for the effective Lagrangians of $D$-meson, charmed, and light baryons from several decay processes of $\Lambda_{c}$ and $\Sigma_{c}$ baryons.
We first calculated the decay widths for the processes $\Lambda_{c} \to D N$, $\Lambda_{c} \to D^{*}N$, $\Sigma_{c} \to D\Delta$, and $\Sigma_{c} \to D^{*}\Delta$ from effective Lagrangian method and quark model picture with the $^{3}P_{0}$ model, and then compared the decay widths from the two models to fix the strength parameter $\lambda$, where the coupling constants $f_{D^{*}\Lambda_{c}N}$ and $h_{D^{*}\Lambda_{c}N}$ from literatures are used as inputs. By utilizing the obtained value of $\lambda$, the coupling constants $g_{D\Lambda_{c}N}$, $g_{D\Sigma_{c}\Delta}$, and $f_{D^{*}\Sigma_{c}\Delta}$ are estimated from the decay widths of the processes $\Lambda_{c} \to D N$, $\Sigma_{c} \to D\Delta$, and $\Sigma_{c} \to D^{*}\Delta$ near threshold. It turns out that the expressions for the coupling constants are independent of the initial masses when one considers decay processes near threshold.

It is found that the coupling constants $g^{(P)}_{D\Lambda_{c}N}$ and $g^{(A)}_{D\Lambda_{c}N}$ derived in this study are consistent with those in the cited literatures.
The estimated coupling constants  $g^{(P)}_{D\Lambda_{c}N}$, $g^{(A)}_{D\Lambda_{c}N}$, $g_{D\Sigma_{c}\Delta}$, and $f_{D^{*}\Sigma_{c}\Delta}$ are expected to be useful for further studies of charmed hadron production.

\section{Acknowledgement}
\label{sec-5}
This work was supported by (i) Suranaree University of Technology (SUT), (ii) Thailand Science Research and Innovation (TSRI), and (iii) National Science Research and Innovation Fund (NSRF), project no. 160355.
TS and YY acknowledge support from Thailand Science Research and Innovation and Suranaree University of Technology through the Royal Golden Jubilee Ph.D. Program (Grant No. PHD/0041/2555).

\end{document}